# Temperature and Spatial Dependence of the Superconducting and Pseudogap of NdFeAsO$_{0.86}$F$_{0.14}$

M. H. Pan,[1][*] X. B. He,[2] G. R. Li,[2] J. F. Wendelken,[1] R. Jin,[3] A. S. Sefat,[3] M. A. McGuire,[3] B. C. Sales,[3] D. Mandrus,[3] and E. W. Plummer[2]

[1]*Center for Nanophase Material Sciences, Oak Ridge National Laboratory, Oak Ridge, TN*

[2]*Physics Department and Astronomy, University Of Tennessee, Knoxville, TN*

[3]*Materials Science and Technology Division, Oak Ridge National Laboratory, Oak Ridge, TN*

**The discovery of superconductivity with a critical temperature exceeding 55 K in the iron-oxypnictides and related compounds has quite suddenly given the community a new set of materials[1–5]—breaking the *tyranny of copper*. This new class of materials raises fundamental questions related to the origin of the electron pairing in the superconducting state and to the similarity to superconductivity in the cuprates. Here, we report spatially resolved measurements using scanning tunneling microscopy/spectroscopy (STM/STS) of the newly discovered iron-based layered superconductor NdFeAsO$_{0.86}$F$_{0.14}$ ($T_c$ = 48 K) as a function of temperature. The tunneling spectra at 17 K show a suppression of spectral intensity within ±10 meV, indicative of the opening of the superconducting gap (SG). Below $T_c$, the sample exhibits two characteristic gaps—a large one (Δ ~ 18 meV) and a small one (Δ ~ 9 meV)—existing in different spatial locations. Both gaps are closed above $T_c$ at the bulk $T_c$, but only the small gap can be fitted with a superconducting gap function. This gap displays a BCS-like order parameter. Above $T_c$, at the same location where the small gap was observed, a pseudogap (PG) opens abruptly at a temperature just above $T_c$**

---

[*] Present address: PO Box 2008, Oak Ridge National Laboratory, Oak Ridge, TN 37831-6487

**and closes at 120 K. In contrast to the cuprates, the SG and PG have competing order parameters.**

With the discovery of high-temperature superconductivity in the iron-oxypnictides and related materials, it now may be possible to study high-temperature superconductivity and its relation to magnetism in a wide range of magnetic element-based materials. The prototypical parent (nonsuperconducting) compounds such as LaFeAsO and CeFeAsO are metallic with a structural distortion below ≈150–160 K that changes the symmetry from tetragonal to orthorhombic at low temperatures. At ≈130–140 K, a long-range spin density wave (SDW)-type antiferromagnetic order with a small moment sets in. Doping the system with F suppresses both the magnetic order and the structural distortion with a concomitant onset of superconductivity.[6] A critical issue is the nature of superconductivity in these materials, especially in comparison with the cuprates—specifically the pairing symmetry and orbital dependence of the SG, the nature of the superconducting order parameter, and the existence of a PG state above $T_c$. STM and STS, as a function of temperature, provide the only way to probe the spatial dependence of these gaps.

Polycrystalline samples with $NdFeAsO_{0.86}F_{0.14}$ nominal composition were prepared by a standard solid-state synthesis method similar to that reported by Sefat et al.[7] The $T_c$ (midpoint temperature = 48 K) and composition were determined by methods described previously.[7] The experiment was carried out using a homemade variable-temperature scanning tunneling microscope in an ultrahigh vacuum chamber (base pressure $<3 \times 10^{-11}$ Torr). This instrument allows simultaneous mapping of atomic-scale topography and differential conductance spectroscopy, proportional to the energy-dependent local density of states (LDOS) of the sample. All STS spectra presented here were obtained at sample biases ranging from +10 to +100 mV with a set point tunneling current of 100 pA. The samples were scraped in ultrahigh vacuum to obtain a clean and fresh surface for STM measurements. Low-temperature topographic scans by STM show a flat surface with corrugation less than 1 nm as displayed in a typical STM constant current image shown in Fig 1(a).  As might be expected, this scraped surface does not display an ordered atomically resolved image.  However, it is still possible to position the tip in one crystalline grain and to measure the local spectroscopy.

To investigate the spatial dependence, a series of point spectra ($dI/dV$) on a 20 × 20 grid within the same area of the topographic image were measured.  Figure 1(b) shows the STS spectra at 17 K measured at the locations of the blue crosses shown in Fig. 1(a).  Careful analysis of the data indicates that the entire 400 spectra fall into two categories.  One set has a pronounced gap state with spectra exhibiting a V-shape [Fig. 1(c) and red curve], while the other set exhibits a more U-shaped character with a flatter bottom [green curve].  Nowhere on the sample surface is there evidence for the coexistence of two gaps, suggesting that multiple gaps observed below $T_c$ are a result of inhomogeneities and spatial averaging.  The V-shaped gap could be fit with a superconducting gap equation whereas the U-shaped gap could not; therefore, the rest of this paper will focus on the V-shaped gap.  However, at $T_c$, the spectra seem to be featureless.

Recent angle resolved photoemission (ARPES) work revealed three Fermi surface (FS) sheets in $Ba_{0.6}K_{0.4}Fe_2As_2$ single crystals.[8,9]  The investigators found nearly isotropic and nodeless SGs of different values opening up on the different FS sheets.  Because PES measures the properties averaged over a range of several millimeters, the inhomogeneous gap distribution on the nanoscale observed here would result in the observation of multigaps in ARPES.

To understand the nature of superconductivity in these materials, it is essential to examine the temperature dependence of tunneling spectra across $T_c$.  For such an inhomogeneous system in which electronic states and the superconducting energy gap vary on the nanometer scale, it is necessary to track specific locations from low temperatures to temperatures above $T_c$.[10]  The measurements shown in Fig. 2 were acquired by keeping the tip at the location with a V-shaped gap as the temperature was changed. The temperature evolution of the gap has been measured by elevating the sample temperature from 17 K (far below $T_c$) to 150 K (far above $T_c$). These results are summarized in Fig. 2(a) and Fig. 2(b) where the normalized spectra are presented. All the measured normalized differential conductance curves show a pronounced symmetric structure within ±100 mV. As the temperature increases, the dip at zero bias is reduced, almost vanishing at about 48.2 K. With further increases of the sample temperatures, a new gap-like feature appears in the STS spectra with a larger gap energy than seen in the SG. This gap feature, which appears only above $T_c$ and has a finite density of states at zero bias, is assigned to a PG state. Such a PG feature has been reported in PES measurements.[11] As the sample temperature is increased, the intensity of this PG slowly decreases while maintaining its identical shape. When the temperature is above 120 K, the whole spectrum becomes a gapless flat curve again.

The normalized $dI/dV$ spectrum can be numerically fitted to obtain the SG size. We used either an *s*- or *d*-wave gap function employing the Dynes function[12] which is convoluted with a Gaussian to take into account the instrumental resolution (see supplementary information). All the data reported were measured with a 1-mV root-mean-square (RMS) modulation, therefore blurring our energy resolution by approximately 2.8 meV. An experiment performed at a temperature of 17.4 K gave a temperature broadening of $\sim k_B T = 1.49$ meV; therefore, in the calculation, a total instrument broadening at this temperature was ~5 meV. From the fit in 17 K [Fig. 3(a)], the value of Δ is estimated to be 9.2 meV. Figure 3(a) shows that at this temperature, both *s*- and *d*-wave fitting curves conform to the data. This gap value (Δ ~ 9.2 meV) is close to the results reported.[13] The reduced gap value $2\Delta/k_B T_c$, estimated with the obtained gap value, is about 4.5, close to what is expected from BCS theory. We fitted the data above $T_c$ using the Dynes function for the PG, as has been done in the past for the cuprates.[14] Notice that there are no sharp superconducting coherence peaks, suggesting that our samples are

"dirty superconductors." When the spectra are fit (see supplementary information) with *s*- or *d*-wave gap functions, a very short inverted lifetime of $\Gamma \sim 3$ meV is obtained. This is ~1000 times larger than that seen in conventional superconductors and about the same as that observed for some cuprate samples at higher temperatures.[15] It is possible that the large inverted lifetime is a result of the polycrystalline nature of the sample.

The temperature dependence of both the PG and SG obtained from our fitting is shown in Fig. 3(c). For T< $T_c$, the measured temperature dependence of $\Delta$ is reproduced quite well by BCS theory (shown by the red curve), except at temperatures close to $T_c$. The measured gap never goes to zero, but reaches a minimum value ~ 3 meV before growing into the PG. This undoubtedly is a signature of the competition between the PG and SG. Figure 2(b) shows that as you increase the temperature above $T_c$, a new gap feature appears in the spectra. When we fitted this gap [supplementary information], we obtained a $\Delta \sim 14$ meV, which is larger than the SG. Fig. 3(c) shows that the PG turns on abruptly as the SG vanishes and displays a very weak dependence upon the temperature until it drops to ~2–3 meV at 120 K. Our result, where the PG only appears above $T_c$ and disappears far above $T_c$, is quite consistent with recent PES data[11] but totally in conflict with the Andreev reflection experiments on $SmFeAsO_{0.85}F_{0.15}$.[16]

The temperature dependence of the PG in this sample is markedly different from the behavior in the high-$T_c$ cuprate superconductors. STS experiments[17] in cuprate superconductors suggested that the SG smoothly connects with the PG at $T_c$, leading to a picture where the order parameter amplitude persists up to $T^*$, above $T_c$, while the macroscopic phase coherence sets in only below $T_c$.[18,19] Our results indicate that the PG order parameter disappears at $T_c$, implying that the PG state is a competing order, in contrast with the cuprate case where the PG helps electron pairing and creates a smooth connection of the SG and PG order parameters. The competition of these two order parameters may be the cause for a slight deviation in the SG order parameter very close to $T_c$. Elucidation of the detailed behavior of the SG and PG near $T_c$ will be crucially important.

In these new materials, the origin of the PG is still a mystery. However, an important clue emerges from neutron scattering data. The disappearance of the static antiferromagnetic order and lattice distortion present in the parent compounds in the doped superconducting materials [LaFeAsO$_{0.92}$F$_{0.8}$ ($T_c$ = 26 K)] suggests that doping the system with fluorine suppresses both the magnetic order and structural distortion in favor of superconductivity. It is reasonable to assume that the PG is related to spin fluctuations in the doped materials.

We conclude by showing that even for these polycrystalline samples, Fourier transformation of the conductance map can reveal k-space information. Spatially resolved differential tunneling conductance $dI/dV(r)$ maps are powerful tools for determining atomic-scale spatial rearrangements of electronic structure. Figure 4(a) displays such a $dI/dV$ map measured in the same field of view as Fig. 1(a) at +20 meV. The Fourier transform shown in Fig. 4(b) and the line scan in Fig. 4(c) show one primary **q**-vector at **q** ~ ± 0.18 ($\pi$/a). The absence of scattering waves in differential conductance maps excludes FS nesting as the explanation. This observation raises the possibility that a striped structure with quasi-periodic length of about a/0.18 ~ 5a exists.

In summary, we are the first to report spatial- and temperature-dependent STM/STS measurements on the FeAs superconductor NdFeAsO$_{0.86}$F$_{0.14}$. Two distinctly different gaps were observed in the tunneling spectrum at 17 K at different places on the sample. The SGs close at $T_c$, displaying a BCS order parameter. By fitting with the Dynes function, we obtained a reasonable gap value ($\Delta$ ~ 9.2 meV), which is quite consistent with other reported results. A PG with a finite density of states (DOS) at the Fermi level (EF) opens at temperatures above $T_c$ and closes at 120 K.   The PG and SG states appear to have competing order parameters, quite different from the cuprates, $MgB_2$, or conventional superconductors.

**Supplementary Information** is linked to the online version of the paper at www.nature.com/nature.

**Acknowledgements**   This research was supported in part (MHP, JFW) by the Laboratory Directed Research and Development Program at Oak Ridge National Laboratory (ORNL), managed by UT-Battelle, LLC for the U. S. Department of Energy (DOE), and in part (RJ, ASS, MAM, BCS, DM) by the Division of Materials Sciences and Engineering (DMS&E), U. S. DOE at ORNL. EWP, XH and GL would like to acknowledge support from the National Science Foundation (NSF) and DOE (DMS&E) through NSF-DMR-0451163 and also the support from The University of Tennessee SARIF program.

**Author Information** Reprints and permissions information are available at www.nature.com/reprints. Correspondence and requests for materials should be addressed to E.W.P.(eplummer@utk.edu).

**Figure Legends:**

**Fig. 1.  Topographic STM image and the two kinds of corresponding spatially resolved spectra.**  (a) Constant current image shows the surface corrugation of an NdFeAsO$_{0.86}$F$_{0.14}$ sample within the 15 × 15 nm$^2$ scan range at 17 K.   The bias is 100 mV, and the current set point is 100 pA.  (b) Set of $dI/dV$ spectra taken at spatial locations of the blue crosses shown in (a).  The U-shaped (V-shaped) curve is highlighted by green (red) color.  (c) Typical $dI/dV$ spectrum chosen for carrying out the temperature-dependent measurements.  All $dI/dV$ data were taken using a lock-in amplifier at 50 mV (200 pA) with 1-mV RMS modulation.  The modulating frequency was 982 Hz.

**Fig. 2. Evolution of gaps with temperature.** Normalized dI/dV spectra for the FeAs superconductor NdFeAsO$_{0.86}$F$_{0.14}$. (a) Below $T_c$ and (b) above Tc showing the evolutions of SG and PG, respectively. Normalization of the spectra was taken by dividing the dI/dV data with simultaneously recorded I/V data.

**Fig. 3. Temperature dependence of gaps.** (a) and (b) The normalized dI/dV spectra at 17 and 42.2 K with experimental data indicated by black circles; fits to the experimental curve with the use of a thermally broadened s-wave (d-wave) BCS model are depicted as red (blue) lines. (b) Temperature dependence of the gap value (open circles) obtained from the fit with the Dynes function. The red curve is the BCS theory prediction. The error bar shown in here is due to the fitting.

**Fig. 4. Differential conductance mapping.** (a) Images of differential conductance dI/dV(r, E) at E = 20 meV in the upper part of the same field of view as Fig. 1(a); the imaging size is 4.5 × 11 nm$^2$. The junction resistance was set to 250 MΩ at V = 20 mV. (b) Fourier transforms of the dI/dV map in Fig 4(a). Obviously, it contains a dominant **q**-vector. (c) The magnitude of the Fourier transform is plotted along the direction of the **q**-vector, which shows the vector length to be about ±0.18 (π/a).

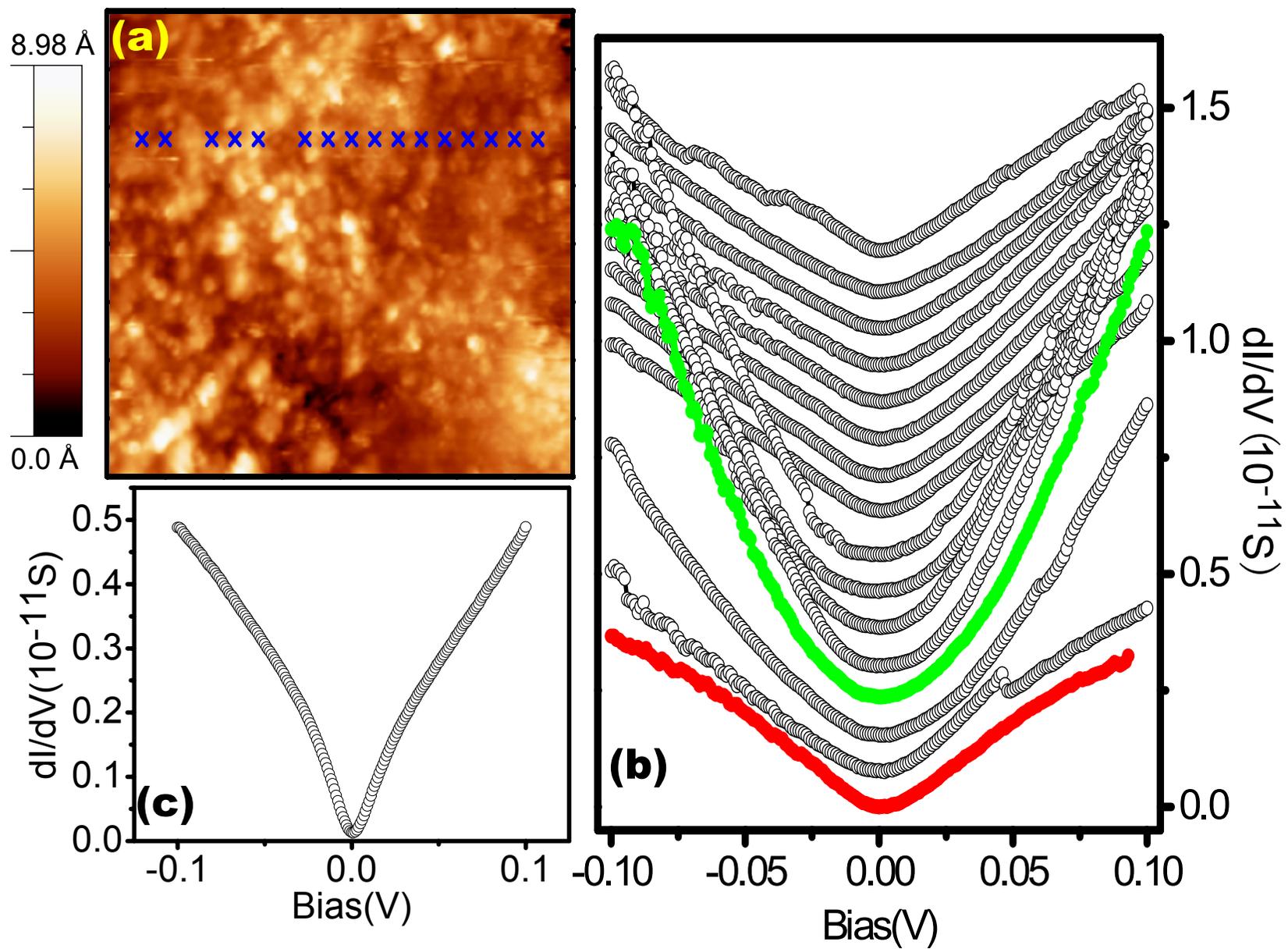

Figure 1, M.H.Pan, *et al*

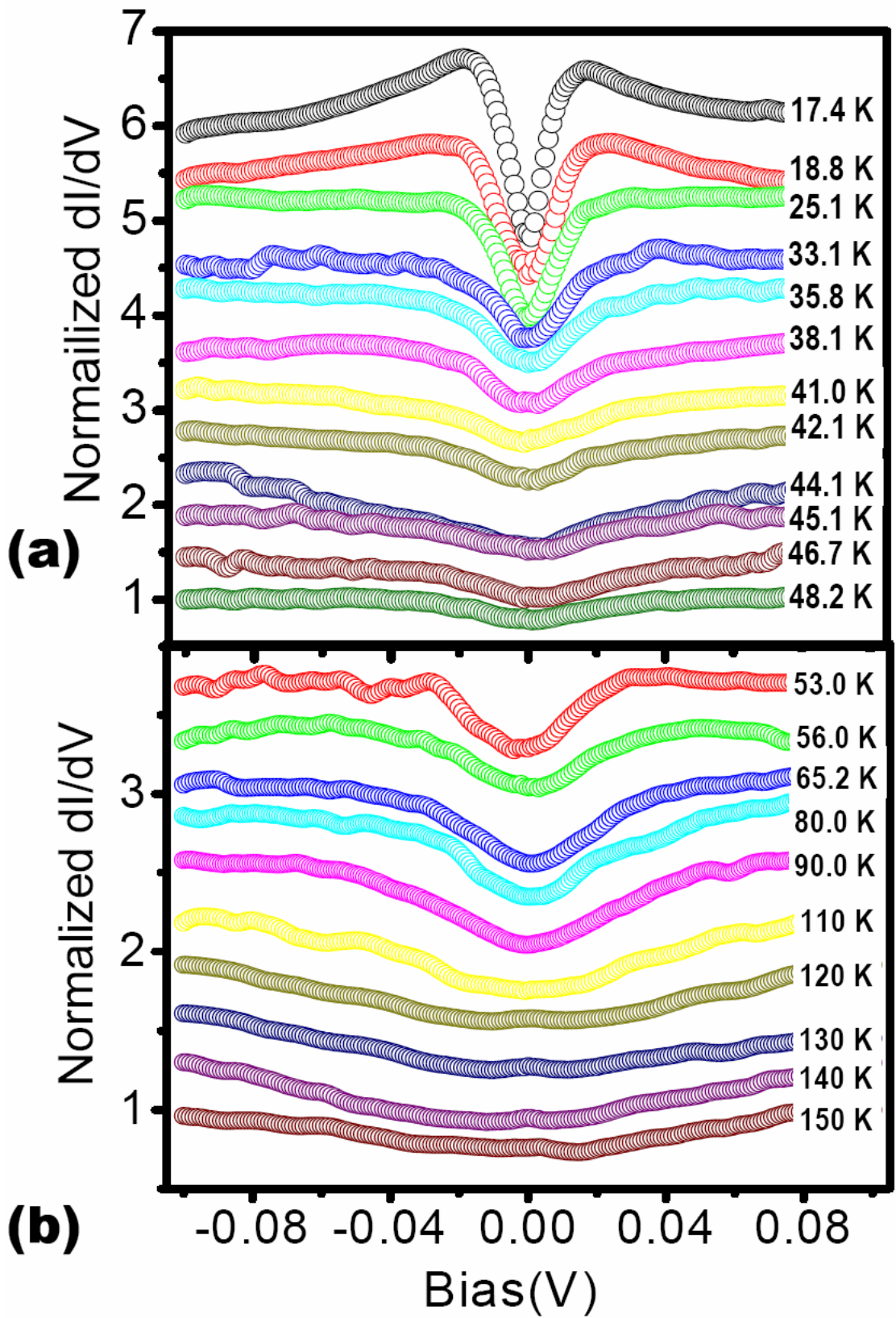

Figure 2, M. H. Pan *et al*

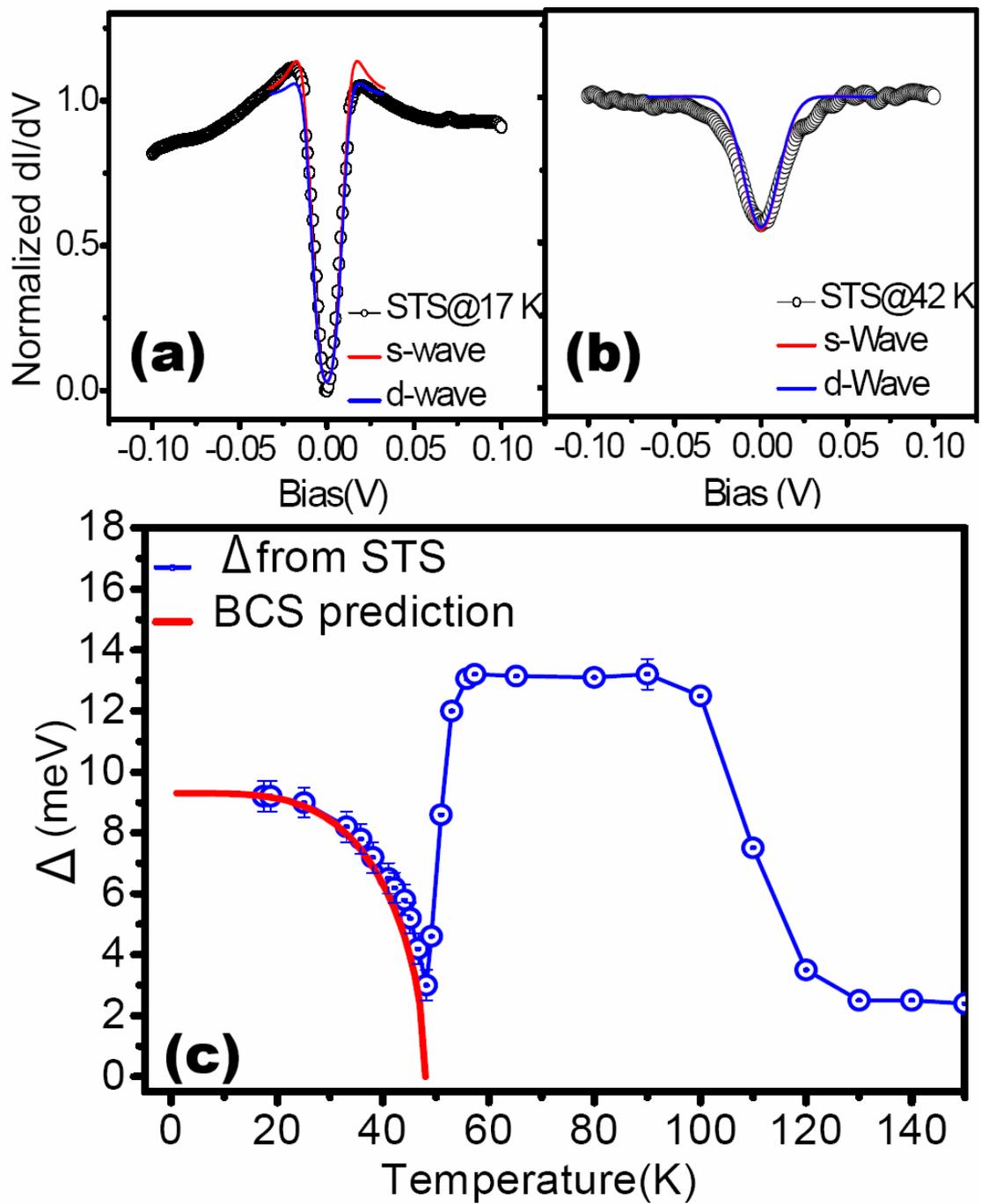

Figure 3, M. H. Pan *et al*

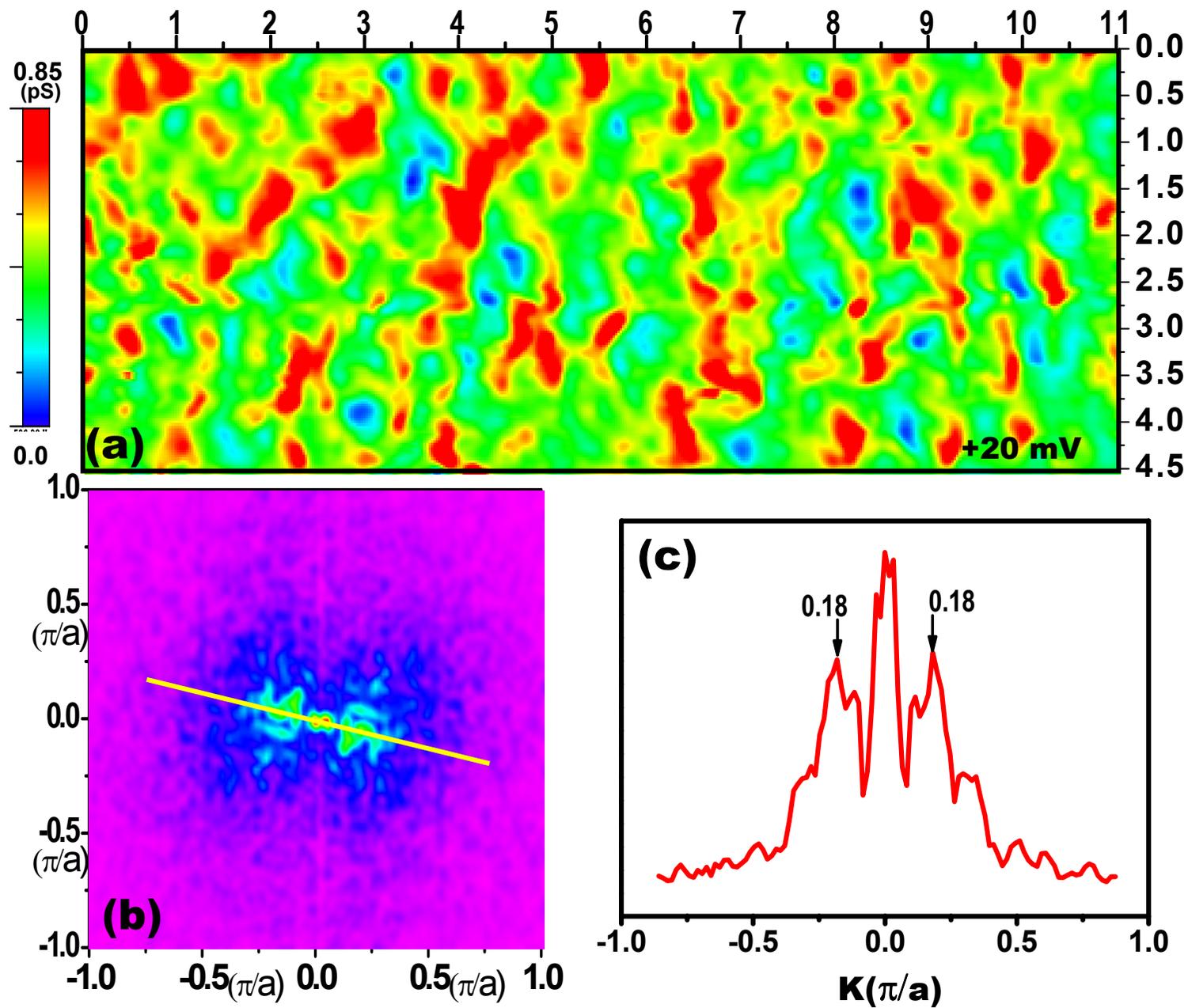

Figure 4, M.H.Pan, *et al*

Supporting online material for

# Temperature and Spatial Dependence of the Superconducting and Pseudogap of NdFeAsO$_{0.86}$F$_{0.14}$

M. H. Pan, X. B. He, G. R. Li, J. F. Wendelken, R. Jin, A. S. Sefat, M. A. McGuire, B. C. Sales, D. Mandrus, and E. W. Plummer

**1. Fitting procedure of superconducting gaps below Tc**

All of the gap data have been fitted with both an s- or d-wave gap function employing the Dynes function [12] which is convoluted with a Gaussian to take into account the instrumental resolution.

**s-wave Dynes function:**

$$G_{SN} = \frac{dI_{SN}}{dV} = \frac{2\pi e}{\eta}|\mu|^2 g_1(0)g_2(0)\{\int_{-\infty}^{\infty} \frac{|\varepsilon - i\Gamma|}{\sqrt{(\varepsilon - i\Gamma)^2 - \Delta^2}} \times \frac{\beta e^{\beta(\varepsilon + eV)}}{(1+\beta e^{\beta(\varepsilon + eV)})^2} d\varepsilon\}$$

**d-wave Dynes function:**

$$G_{SN} = \frac{dI_{SN}}{dV} = \frac{2\pi e}{\eta}|\mu|^2 g_1(0)g_2(0)\{\int_{-\infty}^{\infty} \frac{|\varepsilon - i\Gamma|}{\sqrt{(\varepsilon - i\Gamma)^2 - \Delta^2 \cos^2 2\theta}} \times \frac{\beta e^{\beta(\varepsilon + eV)}}{(1+\beta e^{\beta(\varepsilon + eV)})^2} d\varepsilon\}$$

All the data reported were measured with a 1-mV root-mean-square (RMS) modulation, therefore blurring our energy resolution by approximately 2.8 meV. An experiment performed at a temperature of 17.4 K gave an energy broadening of ~ $4k_BT$ = 5.8 meV of the factor of 4 originates from the fact that the Fermi function smears by approximately $K_BT$ on either side of $E_F$, and the other factor of 2 because we have thermal broadening in both the initial and final states. Therefore, in the calculation, a total instrument broadening at this temperature was ~6.4 meV. From this fit [Fig. 3(a)], the value of Δ is estimated to be 9.2 meV, with a broadening or inverse lifetime Γ of 3.0 meV.

The fit using the s-wave functions as a function of temperature is shown in Figure S1 for T < $T_c$. Since s-wave and d-wave give similar fitting results, here we just show the s-wave fitting at each temperature point.

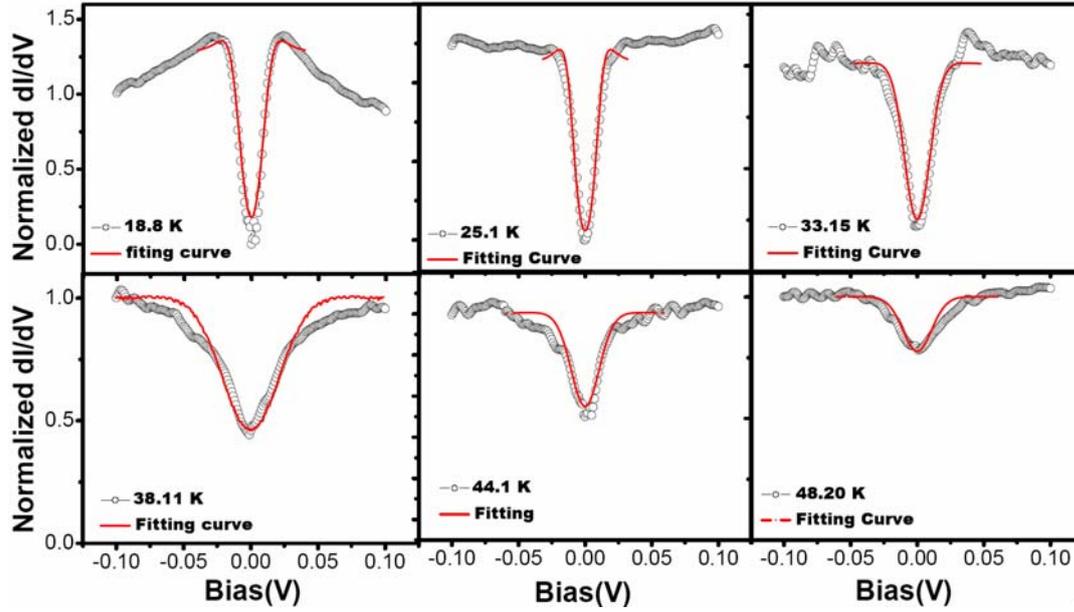

Figure S1: Some experimental dI/dV spectra at different sample temperatures (black hollow dots), and the s-wave fitting results (red curves)

When the temperature is changed, the effect of temperature broadening increases linearly with temperature. The total instrumental resolution includes the contributions from both thermal broadening and bias modulation. Also, the inverse lifetime $\Gamma$ will increase with temperature. The temperature effect on the inverse lifetime has been discussed in detail by Abhay N. Pasupathy et al.[15] They found that the average lifetime broadening dramatically increases with the temperature in $Bi_2Sr_2CaCu_2O_{8+\delta}$, because the sample loses long-range phase coherence. In our fitting for different temperature spectra, we find a similar temperature dependence of the inverse lifetime $\Gamma$. The temperature dependences of the two fitting parameters, instrumental resolution and inverted lifetime are shown in figure S2.

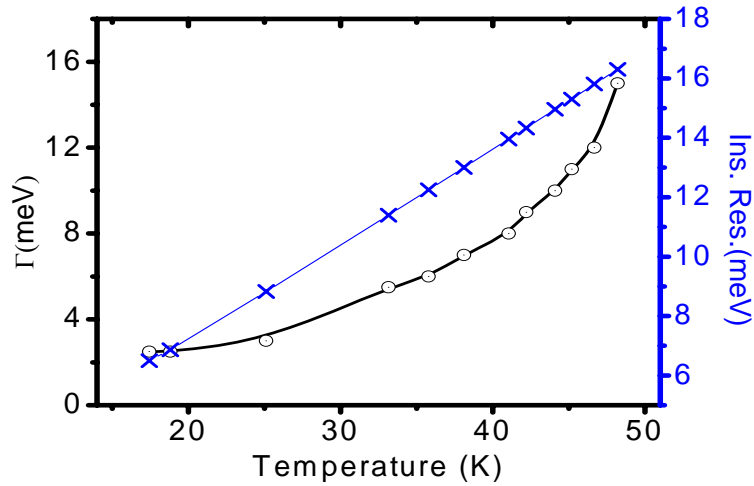

Figure S2: Extracted lifetime broadening and instrumental resolutions plotted as a function of temperature. Black hollow dots: The average lifetime broadening as a function of temperature. Blue cross points: Instrumental resolution.

## 2. Fitting for the pseudogap above Tc

We do not know the origin of the pseudogap but if it is due to a partial pairing state of electrons, we can still use the Dynes function to fit the data above Tc. Figure S3 shows a typical fitting for spectra at 90 K, which give a pseudogap value about 26 meV ($\Delta \sim 13$ meV). For the pseudogap above Tc, the fitting becomes more insensitive for a $\Gamma$ value larger than 15 meV.

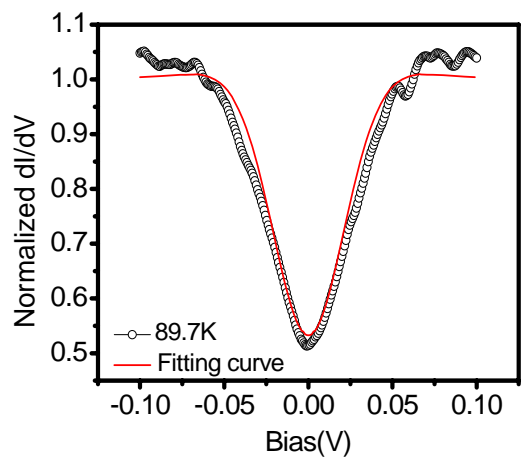

Figure S3: A example shows the fit for pseudogap at 90 K by s-wave Dynes function.

### 3. Difficulty in fitting the U-shaped spectra at 17 K

When we tried to fit all our dI/dV spectra for the U-shaped gap below $T_c$ using the Dynes function, we found it to be impossible. The main reason is that the normalized curve has a large opening without the flat bottom, shown in Figure S4. A large opening between the gap edges means a large gap value, but a large superconducting gap definitely will produce a large flat gap bottom, as shown by the red fitting curve (given $\Delta \sim 18$ meV). Here, we use the same $\Gamma$ gamma value as for the V-shaped one at 17K for fitting the U-shaped curve.

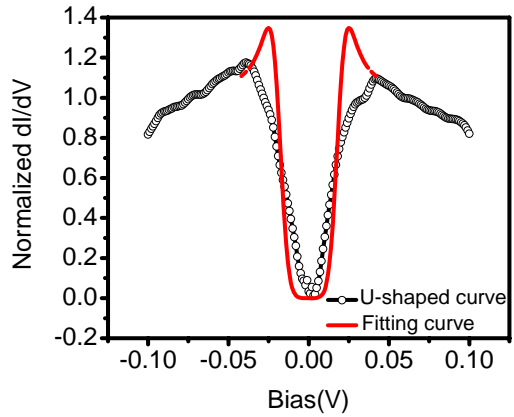

Figure S4: A example shows the fit for U-shaped spectra at 17 K by s-wave Dynes function.